\documentclass[aps,prd,preprint,showpacs,superscriptaddress]{revtex4}

\usepackage{graphics}

\usepackage{enumerate}
\usepackage{amsmath,amsthm,amssymb}
\usepackage{slashed}
\usepackage{array}
\usepackage{latexsym}

\usepackage{graphicx}

\usepackage{pstricks}

\usepackage{epsfig}

\begin{document}

\title{$SU(3)_c\otimes SU(3)_L\otimes U(1)_X$ models with four families}

\author{Richard H. Benavides}

\affiliation{Instituto de F\'\i sica, Universidad de Antioquia,
A.A. 1226, Medell\'\i n, Colombia.}
\affiliation{Instituto Tecnol\'ogico Metropolitano, Facultad de Ciencias, Medell\'in, Colombia}
\author{William A. Ponce}
\affiliation{Instituto de F\'\i sica, Universidad de Antioquia,
A.A. 1226, Medell\'\i n, Colombia.}
\author{Yithsbey Giraldo }
\affiliation{Instituto de F\'\i sica, Universidad de Antioquia,
A.A. 1226, Medell\'\i n, Colombia.}
\affiliation{Departamento de F\'isica, Universidad de Nari\~no, A.A. 1175, Pasto, Colombia}

\date{\today}

\begin{abstract}
In the context of the local gauge group $SU(3)_c\otimes SU(3)_L\otimes U(1)_X$, we look for possible four family models, where all the particles carry ordinary electric charges. Thirteen different anomaly-free fermion structures emerge, out of which only two are realistic. For the simplest physical structure, we calculate the charged and neutral weak currents and the tree-level Fermion masses. We also look for new sources of flavor changing neutral currents in the quark sector, in connection with the upcoming experimental results at the Large Hadron Collider.
\end{abstract}

\maketitle

\section{\label{sec:sec1}Introduction}
The standard model (SM) based on the local gauge group $SU(3)_c\otimes SU(2)_L\otimes U(1)_Y$ \cite{jf} has been very successful so far, in the sense that its predictions are in good agreement with present experimental results. However, the SM fails short in explaining things as: hierarchical charged fermion masses, fermion mixing angles, charge quantization, strong CP violation, replication of families, neutrino masses and oscillations~\cite{RN}, or the matter-antimatter asymmetry of the universe. It also does not contain a good candidate for the dark mater component of the universe~\cite{dark}.

The SM does not predict the number $N$ of Fermion families existing in nature; the only restriction, $N\leq 8$ comes from the asymptotic freedom of $SU(3)_c$ also known as quantum-cromo dynamics or QCD~\cite{jf}. Experimental results at the CERN-LEP early in the 1990's, implied the existence of at least three families each one having a neutral lepton with a mass less than half the mass of the neutral $Z^0$ gauge boson; result interpreted at the beginning as an exact value for the total number of families in nature, which is not quite correct. As a matter of fact, the LEP data does not exclude the existence of extra families having heavy neutrinos.

Determining the number of fermion families is a key goal of the upcoming experiments at the large hadron-Collider (LHC) \cite{hol}, and further at the ILC \cite{Ciftci}. In principle, the existence of new heavy quark flavors and their mixing with the ordinary ones is possible, due to the fact that the uncertainties on the measured Cabibbo-Kobayashi-Maskawa (CKM) matrix elements \cite{yao} left an open door for this, with a fourth family of quarks $(t',b')$ and their mixing with the other three families, not ruled out yet. Experiments at the Tevatron have already constrained the masses of a fourth family of quarks to be $m_{t'}> 311$ GeV and $m_{b'}> 338$ GeV \cite{Aatonen}. 

Theoretical constrains on the masses of the fourth family fermions $t'$,$b'$,$\tau'$, and $\nu'_\tau$, are obtained from the one loop contributions to the corrections parameter S and T \cite{Gates}. In contrast to some previous claims, a fourth sequential family is not in conflict with precision measurements~\cite{plehn} of the electroweak parameters. Remarkably enough, fourth family fermions with masses around $550$ GeV would couple strongly to the Goldstone Bosons of the electroweak symmetry breaking \cite{hol}, producing condensates which mimic in some way the effect of the Higgs scalars, joining in this way the issue of the flavor problem with the, until now obscure spontaneous symmetry breaking mechanism.

It is clear thus that, there is not experimental or phenomenological evidence which exclude the existence of a fourth family with a heavy neutrino. Indeed, the recent electroweak precision data are equally consistent with the presence of three or four families \cite{He}, whereas the four family scenario is favored if the Higgs mass is heavier than $200$ GeV \cite{Kribs}.

Among the several extensions of the SM proposed so far, the models based on the $SU(3)_c\otimes SU(3)_L\otimes U(1)_X$  local gauge group (called here after 3-3-1 for short) are outstanding because they can ameliorate in a natural way, several short-comings of the SM \cite{pf, vl, ozer, sher, pfs, pgs}, (see section II).

One common belief is that consistent 3-3-1 models can exist only for 3,6,9... families (multiple of three which is the number of colors)\cite{pf, vl, ozer}. This result which is valid for a particular way of canceling the gauge anomalies, is not true in general as we are going to see. For example, $E_6$ \cite{gursey} as an anomaly free grand unified theory has a 3-3-1 sub-algebra, which is anomaly free family by family (as in the SM), so it can be the source of an $SU(3)_c\otimes SU(3)_L\otimes U(1)_X$ model with as many families as wished (the 3-3-1 one family model, subgroup of $E_6$, was already analyzed in Ref.\cite{luis}).

In this paper we are going to study models based on the local gauge group $SU(3)_c\otimes SU(3)_L\otimes U(1)_X$ with four families. In particular, we are going to find all the possible 3-3-1 four family anomaly-free structures  without exotic electric charges and find the realistic ones, picking up the simplest one of them in order to see some of its implication for the up-coming experimental results at the LHC.

The paper is organized as follows: in Sec.~\ref{sec:sec2} we find all the possible 3-3-1 four family structures without exotic electric charges, find the realistic ones, and present their common gauge boson and Higgs scalar sectors; in Sec.~\ref{sec:sec3} we introduce the particular model chosen for our analysis and calculate its electroweak currents (charged and neutral); in Sec.~\ref{sec:sec4} we analyze the fermion masses for the particular model under consideration and set our notation; in Sec.~\ref{sec:sec5} we obtain the constraints coming from flavor changing neutral currents (FCNC) and calculate the maximum mixing allowed of the new quarks with the known ones, and finally, Sec.~\ref{sec:sec6} contains our conclusions.


\section{\label{sec:sec2}3-3-1 models without exotic electric charges} 
Popular extensions of the SM are based on the local gauge group  $SU(3)_c\otimes SU(3)_L\otimes U(1)_X$. The several possible fermion and scalar structures enlarge the SM in its gauge, scalar, and fermion sectors. Let us mention some outstanding features of 3-3-1 models:

\begin{itemize}
\item 3-3-1 models free of anomalies can be constructed for one, two, three, four, five and  more families \cite{pfs}. 
\item A Peccei-Quinn chiral symmetry can be implemented easily~\cite{pq, pal} for some 3-3-1 models.
\item For some models with three families, one quark family has different quantum numbers than the other two, fact that may be used to explain the heavy top quark mass~\cite{ff, canal}.
\item The scalar sector includes several good candidates for dark matter~\cite{sanchez}. 
\item The lepton content is suitable for explaining some neutrino properties~\cite{kita}. 
\item The hierarchy in the Yukawa coupling constants can be avoided by implementing several universal see-saw mechanisms~\cite{canal, seesaw, dwl}.
\end{itemize}

In Refs.~\cite{sher, pfs, pgs} the classification of 3-3-1 models without exotic electric charges for three families was presented, where  eight different models were founded. In this section we will do a similar study but for 4 families. Our finding is that there are thirteen different anomaly-free four family 3-3-1 structures without exotic electric charges, out of which only a few ones can be realistic.

To start with, let us consider the following six closed sets of chiral fields (closed in the sense that each set includes the antiparticles of each charged particle), where the quantum numbers in parenthesis refer to the $[SU(3)_c,SU(3)_L,U(1)_X]$ representations.

\begin{itemize}
\item $S_1=[(\nu^0_\alpha,\alpha^-,E_\alpha^-);\alpha^+;E_\alpha^+]_L$ with quantum numbers $(1,3,-2/3);(1,1,1)$ and $(1,1,1)$ respectively.
\item $S_2=[(\alpha^-,\nu_\alpha,N_\alpha^0);\alpha^+]_L$ with quantum numbers $(1,3^*,-1/3)$ and $(1,1,1)$ respectively.
\item $S_3=[(d,u,U);u^c;d^c;U^c]_L$ with quantum numbers $(3,3^*,1/3);\; (3^*,1,-2/3);\; (3^*,1,1/3)$ and $(3^*,1,-2/3)$ respectively.
\item $S_4=[(u,d,D);u^c;d^c;D^c]_L$ with quantum numbers $(3,3,0);\; (3^*,1,-2/3);\; (3^*,1,1/3)$ and $(3^*,1,1/3)$ respectively.
\item $S_5=[(e^-,\nu_e,N_1^0);(E^-,N_2^0,N_3^0);(N_4^0, E^+,e^+)]_L$ with quantum numbers $(1,3^*,-1/3)$;$(1,3^*,-1/3)$ and $(1,3^*,2/3)$ respectively.
\item $S_6=[(\nu_e, e^-,E_1^-);(E^+_2,N_1^0,N_2^0);(N_3^0, E^-_2,E_3^-)$;  $e^+; E_1^+; E_3^+]_L$ with quantum numbers $(1,3,-2/3)$; $(1,3,1/3)$; $(1,3,-2/3)$; $(111), (111)$;  and $(111)$ respectively.
\end{itemize}

The former set of fields is exhaustive, in the sense that any other set will include either exotic electric charges,  3-3-1 vector like representations, or the anomaly-free singlet representation (1,1,0) (a kind of right-handed neutrino). The several triangle anomalies for the former six sets of fields are presented in Table I, which in turn allows us to build anomaly-free 3-3-1 models for one, two, three or more families.

\begin{table}[here!]{{TABLE I: Anomalies for 3-3-1 fermion fields structures}}\label{tabl1}

\begin{tabular}{||l|cccccc||}\hline\hline
Anomalies & $S_1$ & $S_2$ & $S_3$ & $S_4$ & $S_5$ & $S_6$ \\ \hline
$[SU(3)_C]^2U(1)_X$ & 0 & 0 & 0 & 0 & 0 & 0 \\
$[SU(3)_L]^2U(1)_X$ & $-2/3$  & $-1/3$ & 1 & 0& 0 & -1\\
$[Grav]^2U(1)_X$ & 0 & 0 & 0 & 0 & 0 & 0 \\
$[U(1)_X]^3$ & 10/9 & 8/9 & $-12/9$ & $-6/9$& 6/9& 12/9 \\
$[SU(3)_L]^3$ & 1 & $-1$ & $-3$ & 3 & $-3$ & 3\\
\hline\hline
\end{tabular} 
\end{table}

\subsection{\label{sec:sec21}Four family models}
From the former table we can construct the following thirteen anomaly-free 3-3-1 structures for four families of quarks and leptons:
\begin{itemize}
\item[1]: $4(S_4+S_5)$,
\item[2]: $4(S_3+ S_6)$,
\item[3]: $2(S_1 + S_2+ S_3+ S_4)$,
\item[4]: $2(S_3+S_4+S_5+S_6)$,
\item[5]: $3S_1 + 2S_3+ 2S_4+ S_5$,
\item[6]: $3S_2 + 2S_3+ 2S_4+ S_6$,
\item[7]: $S_1 + S_2+ 2S_3+ 2S_4+S_5+S_6$,
\item[8]: $3S_2+ S_5+ 3S_4+ S_3$,
\item[9]: $3S_4 + 3S_5+ S_3+ S_6$,
\item[10]: $S_1 + S_2+ 2S_5+ 3S_4+S_3$,
\item[11]: $3S_3 + S_4+ S_5+ 3S_6$,
\item[12]: $S_1 + S_2+ 2S_6+ S_4+ 3S_3$,
\item[13]: $3S_1 + S_6+ 3S_3+ S_4$.
\end{itemize}
Let us make some comments:

\noindent
a) Structures 1 and 2 are four family models, carbon copy of the one family anomaly-free gauge structures already studied in Refs. \cite{luis} and \cite{marti} respectively. Model one has 4 up type quarks and 8 down type quarks; model two has 8 up type quarks and 4 down type quarks.\\
b) Structures 3 and 4 are four family models, carbon copy of two gauge models for two families.\\
c) For the five four family structures in models 3,4,5,6, and 7, the number of up type quarks is equal to the number of down type quarks; equal to six.\\
d) The three four family structures 8, 9, and 10 all have five up type quarks and seven down type quarks.\\
e) The three four family structures 11, 12 and 13 all have seven up type quarks and five down type quarks.\\
f) The simplest lepton structure is the third one, with six charged leptons and six Weyl neutral states.

\subsection{\label{sec:sec22}The Gauge boson structure}
All the 3-3-1 local Gauge models without exotic electric charges have the same Gauge boson structure, dictated by the group properties, independent of the number of families in each particular model and of the fermion field content. There are in total 17 gauge Bosons: one
gauge field $B^\mu_X$ associated with $U(1)_X$, the 8 gluon fields $G^\mu_\nu$
associated with $SU(3)_c$ which remain mass-less after the spontaneous breaking of the electroweak 
symmetry, and another 8 gauge fields associated with $SU(3)_L$ that we
write for convenience as \cite{pgs}
\begin{equation}\label{maga}
\sum_{\alpha=1}^8\lambda^\alpha A^\mu_\alpha=\sqrt{2}\left(
\begin{array}{ccc}D^\mu_1 & W^{+\mu} & K^{+\mu} \\ W^{-\mu} & D^\mu_2 &
K^{0\mu} \\ K^{-\mu} & \bar{K}^{0\mu} & D^\mu_3 \end{array}\right), 
\end{equation}
where $D^\mu_1=A_3^\mu/\sqrt{2}+A_8^\mu/\sqrt{6},\;
D^\mu_2=-A_3^\mu/\sqrt{2}+A_8^\mu/\sqrt{6}$, and
$D^\mu_3=-2A_8^\mu/\sqrt{6}$. $\lambda_\alpha, \; \alpha=1,2,...,8$, are the eight
Gell-Mann matrices normalized as $Tr(\lambda^\alpha\lambda^\beta)  
=2\delta_{\alpha\beta}$.

The charge operator associated with the unbroken gauge symmetry $U(1)_Q$ 
is given by:
\begin{equation}\label{chargo}
Q=\frac{\lambda_{3L}}{2}+\frac{\lambda_{8L}}{2\sqrt{3}}+XI_3
\end{equation}
where $I_3=Diag.(1,1,1)$ is the diagonal $3\times 3$ unit matrix, and the 
$X$ values are related to the $U(1)_X$ hypercharge and are fixed by 
anomaly cancellation. 
The sine square of the electroweak mixing angle is given by 
\begin{equation}\label{ewk}
S_W^2=3g_1^2/(3g_3^2+4g_1^2)
\end{equation}
where $g_1$ and $g_3$ are the coupling 
constants of $U(1)_X$ and $SU(3)_L$ respectively, and the photon field is 
given by~\cite{ozer, pgs} 
\begin{equation}\label{foton}
A_0^\mu=S_WA_3^\mu+C_W\left[\frac{T_W}{\sqrt{3}}A_8^\mu + 
\sqrt{(1-T_W^2/3)}B^\mu\right],
\end{equation}
where $C_W$ and $T_W$ are the cosine and tangent of the electroweak mixing 
angle, respectively.

There are two weak neutral currents in the model, associated with the two flavor diagonal neutral gauge weak Bosons, which in the gauge basis can be written as
\begin{eqnarray}\nonumber \label{zzs}
Z_0^\mu&=&C_WA_3^\mu-S_W\left[\frac{T_W}{\sqrt{3}}A_8^\mu + 
\sqrt{(1-T_W^2/3)}B^\mu\right], \\ \label{zetas}
Z_0^{\prime\mu}&=&-\sqrt{(1-T_W^2/3)}A_8^\mu+\frac{T_W}{\sqrt{3}}B^\mu,
\end{eqnarray}
and another electrically neutral current associated with the gauge boson $K^{0\mu}$. In the former expressions 
$Z^\mu_0$ coincides with the weak neutral gauge boson of the SM~\cite{ozer, pgs}. 

Let us emphasize that equations (\ref{maga}-\ref{zzs}) presented here are common to all the 3-3-1 gauge structures without exotic electric charges~\cite{vl, ozer, sher}, independent of the scalar sector, of the number of families, and also of the fermion field content for each particular model.

\subsection{\label{sec:sec23}The scalar sector}
Again, all the 3-3-1 local gauge models without exotic electric charges may have in common the same Higgs scalars structure, independent of the fermion representation we are referring to; and thus, independent of the number of families in the model.

In our analysis we are going to use the set of three Higgs scalars introduced in the original papers~\cite{vl, ozer}  (the economical set consisting only of two Higgs scalars~\cite{pgs} or the enlarged set with four Higgs scalars~\cite{canal} are the other two alternatives).

The set of three scalar fields and their vacuum expectation values (VEV) are:
\begin{subequations}
\label{higg}
\begin{align}\label{hig1}
 \Phi_1(1,3^*,-1/3)&=
\begin{pmatrix}
 \phi_1^-\\
\phi_1^0\\
\phi_1^{\prime 0}
\end{pmatrix},\:\textrm{with VEV:}\:
\langle\Phi_1\rangle=
\frac{1}{\sqrt{2}}
\begin{pmatrix}
 0\\
v_1\\
V_1
\end{pmatrix},\\ \label{hig2}
\Phi_2(1,3^*,-1/3)&=
\begin{pmatrix}
 \phi_2^-\\
\phi_2^0\\
\phi_2^{\prime 0}
\end{pmatrix},\:\textrm{with VEV:}\:
\langle\Phi_2\rangle=
\frac{1}{\sqrt{2}}
\begin{pmatrix}
 0\\
v_2\\
V_2
\end{pmatrix},\\ \label{hig3}
\Phi_3(1,3^*,2/3)&=
\begin{pmatrix}
 \phi_3^0\\
\phi_3^+\\
\phi_3^{\prime+}
\end{pmatrix},\:\textrm{with VEV:}\:
\langle\Phi_3\rangle=
\frac{1}{\sqrt{2}}
\begin{pmatrix}
v_3\\
0\\
0
\end{pmatrix},
\end{align}
\end{subequations}
with the hierarchy $v_1\sim v_2\sim v_3\sim 10^2$ GeV $<< V_1\sim V_2\sim 1$ TeV.

The set of scalars and VEV in Eq.~(\ref{hig1}) properly breaks the 
$SU(3)_c\otimes SU(3)_L\otimes U(1)_X$ symmetry in two steps, 
\begin{eqnarray*}
SU(3)_c\otimes SU(3)_L\otimes U(1)_X &\stackrel{(V_A)}{\longrightarrow}& \\
SU(3)_c\otimes SU(2)_L\otimes U(1)_Y &\stackrel{(v_B)}{\longrightarrow}& SU(3)_c\otimes U(1)_Q,
\end{eqnarray*}
for $A=1,2;\;\;B=1,2,3$. This allow for the matching conditions $g_2=g_3$ and 
\begin{equation}\label{mc}
\frac{1}{g^{2}_Y}=\frac{1}{g_1^2}+\frac{1}{3g_2^2},
\end{equation}
where $g_2$ and $g_Y$ are the gauge coupling constants of 
the $SU(2)_L$ and $U(1)_Y$ gauge groups in the SM, respectively. 

This set of three scalar Higgs with the VEV as stated is enough to produce tree-level masses for all the charged fermion fields of any three or four family model.

\subsection{The neutrino sector}
The 13 four family structures defined in Sec.~\ref{sec:sec21} are renormalizable (anomaly-free), but unfortunately, in the form they are presented, most of them are ruled out by neutrino phenomenology. Let us see why:

From the 13 structures, the one with the simplest Fermion content is number 3 which has the lepton fields present in $2S_1+2S_2$, with 6 different charged particles (12 Weyl state) and 6 neutral Weyl states; lepton structure that we may write for convenience as
\begin{subequations}\label{fer31}
\begin{align}
\psi_{eL}&=(\nu_e^0,e^-,E^-_e)_L\sim (1,3,-2/3),\quad e^+_L, E_{eL}^+,\\ \label{fer32}
\psi_{\mu L}&=(\nu_\mu^0,\mu^-,E^-_\mu)_L\sim (1,3,-2/3),\quad \mu^+_L, E_{\mu L}^+,\\ \label{fer33}
\psi_{\tau L}&=(\tau^-,\nu_\tau^0,N^0_\tau)_L\sim (1,3^*,-1/3),\quad \tau^+_L,\\ \label{fer34}
\psi_{\tau^\prime L}&=(\tau^{\prime-},\nu_\tau^{\prime0},N^0_{\tau^\prime})_L\sim (1,3^*,-1/3),\quad 
\tau^{\prime+}_L.
\end{align}
\end{subequations}
Using the scalar Higgs fields and VEV as introduced in (\ref{higg}), the only Yukawa mass terms for the fermion neutral Weyl states of this model are of the form
\begin{eqnarray}\label{yuk3}
{\cal L}^{(3)}_n&=&h_n\epsilon_{ijk}\langle\phi_3^{i*}\rangle\psi_{\tau L}^j\psi_{\tau^\prime L}^k +h.c.\\ \label{yu3f}
&=&h_nv_3(\nu_\tau^0N_{\tau^\prime}^0-N^0_\tau\nu^0_{\tau^\prime})+h.c.,
\end{eqnarray}
which represents two Dirac masses at the electroweak scale, involving four spin 1/2 Weyl states, leaving only room for two mass-less Weyl states which in our notation are $\nu^0_e$ and $\nu^0_\mu$ (they may pick-up masses via quantum corrections), with the inconvenience that, by taking a Yukawa coupling constant $h_n$ very small, will produce four states with masses smaller than $M_Z/2$ instead of the experimentally allowed number of 3. So, this model as stated is ruled out, unless new ingredients are added (as for example an extra neutral Weyl state $N^0_{\tau^{\prime\prime}}\sim (1,1,0)$ followed of a fine tuning of some Yukawa coupling constants, to secure very small masses for three neutrinos). Notice that the use of the economical set of Higgs scalars~\cite{pgs}, or the enlarged set~\cite{canal}, does not solve this problem at all.

A new interesting ingredient appears for structure five which has the Fermion content present in $3S_1+S_5+2(S_3+S_4)$. The lepton content for this particular four family structure can be written as 
\begin{subequations}\label{fer15}
\begin{align}\label{ferq51}
\psi_{eL}&=(\nu_e^0,e^-,E^-_e)_L\sim (1,3,-2/3),\quad e^+_L, E_{eL}^+,\\ \label{ferl52}
\psi_{\mu L}&=(\nu_\mu^0,\mu^-,E^-_\mu)_L\sim (1,3,-2/3),\quad \mu^+_L, E_{\mu L}^+,\\ \label{ferl53}
\psi_{\tau L}&=(\nu_\tau^0,\tau^-,E^-_\tau)_L\sim (1,3,-2/3),\quad \tau^+_L, E_{\tau L}^+,\\ \label{ferl54}
\psi_{\tau^\prime L}&=(\tau^{\prime-}, \nu^0_{\tau^\prime},N^0_1)_L \sim (1,3^*,-1/3), \\ \label{ferl55}
\psi_{1 L}&=(E_2^-,N_2^0,N_3^0)_L\sim (1,3^*,-1/3),\\ \label{fer156}
\psi_{2 L}&=(N_4^0,E_2^+,\tau^{\prime+})_L\sim (1,3^*,2/3).
\end{align}
\end{subequations}
When all the possible Yukawa terms for the Weyl neutral states are included, the following $8\times 8$ mass matrix in the basis $(\nu_e,\nu_\mu,\nu_\tau,\nu_{\tau^\prime},N_1^0, N_2^0,N_3^0,N_4^0)_L$ is obtained
\begin{equation}\label{man5}
\left(
\begin{array}{cccccccc}
0 & 0 & 0 & 0 & 0 & 0 & 0 & M_1 \\
0 & 0 & 0 & 0 & 0 & 0 & 0 & M_2\\
0 & 0 & 0 & 0 & 0 & 0 & 0 & M_3\\
0 & 0 & 0 & 0 & 0 & 0 & a & B_1 \\
0 & 0 & 0 & 0 & 0 & -a & 0 & b_1 \\
0 & 0 & 0 & 0 & -a^* & 0 & 0 & b_2 \\
0 & 0 & 0 & a^* & 0 & 0 & 0 & B_2\\
M_1^* & M_2^* & M_3^* & B_1^* & b_1^* & B_2^* & b_2^* & 0 \\
\end{array}\right),
\end{equation}
where the entries $a,\;b_1,\;b_2$ are Dirac mass terms proportional to the SM mass scale $v_3$, $B_1$ and $B_2$ are Majorana mass terms proportional to the 3-3-1 mass scale $V_A$  and the entries $M_A,\;\; A=1,2,3$ are Majorana mass terms coming from the bare Lagrangian
\[ \psi_{2L}(M_1\psi_{eL}+M_2\psi_{\mu L}+M_3\psi_{\tau L})+ h.c.,\]
which can be as large as the Planck scale. Due to the presence of this last mass entries, the rank of the previous matrix is six, with two eigenvalues equal to zero. Unfortunately, the large bare mass entries can not be used to generate see-saw mechanisms and only, if we allow a discreet symmetry which forbids this bare mass entries, the matrix could have three zero mass states, becoming in this way a realistic one.

Analysis similar to the previous ones have been carried through for the neutrino sector of the thirteen anomaly-free lepton structures enumerated in Sec.~\ref{sec:sec21}. The results are presented in Table~(\ref{table2}).

\begin{table}[here!]{{TABLE IID : Neutrino sectors  }}\label{table2}

\begin{tabular}{|c|c|c|}\hline\hline
Structure & Number of Weyl  & Mas-less  \\
          & neutral states  & states       \\ \hline
1:& 20 & 0  \\
2:& 16 & 0  \\
3:& 6 & 2  \\
4:& 18 & 0  \\
5:& 8 & 2  \\
6:& 10 & 2  \\
7:& 12 & 0  \\
8:& 11 & 1  \\
9:& 19 & 0  \\
10:& 13 & 1  \\
11:& 17 & 0  \\
12:& 11 & 3  \\
13:& 7 & 3  \\
\hline\hline
\end{tabular} 
\end{table}
According to this Table, only structures 12 and 13 survive the natural condition of having 3 tree-level zero mass neutrinos, which may pick up non zero masses via radiative corrections. Some other structures may become realistic if new fields are added, and/or if some Yukawa coupling constants are fine tuned to very small values as mentioned before.


\section{\label{sec:sec3}A four family model} 
In this section we start the study of one of the two four family models which survives in a natural way the precision measurements of the electroweak parameters. In particular, we choose model thirteen with the Fermion  structure given by $3S_1+S_6 +3S_3+S_4$. The Fermion content for this particular four family model is written in the following way
\begin{subequations}\label{fer16}
\begin{align}\label{ferq131}
\psi_{QL}^i&=(d^i,u^i,U^i)_L^T\sim(3,3^*,1/3), \quad d^{ic}_L, u^{ic}_L, U^{ic}_L, \;i=1,2,3\\ \label{q132}
\psi_{QL}^4&=(u^4,d^4,D)_L^T\sim(3,3,0), \quad u^{4c}_L, d^{4c}_L, D^{c}_L,\\ \label{q133}
\psi_{eL}&=(\nu_e^0,e^-,E^-_e)_L\sim (1,3,-2/3),\quad e^+_L, E_{eL}^+,\\ \label{fer132}
\psi_{\mu L}&=(\nu_\mu^0,\mu^-,E^-_\mu)_L\sim (1,3,-2/3),\quad \mu^+_L, E_{\mu L}^+,\\ \label{fer133}
\psi_{\tau L}&=(\nu_\tau^0,\tau^-,E^-_\tau)_L\sim (1,3,-2/3),\quad \tau^+_L, E_{\tau L}^+,\\ \label{fer134}
\psi_{\tau^\prime L}&=(\nu^0_{\tau^\prime}, \tau^{\prime -},E_{\tau^\prime}^-)_L\sim (1,3,-2/3),
\quad \tau^{\prime +}_L, E_{\tau^\prime L}^+\\ \label{fer135}
\psi_{1 L}&=(E_1^+,N_1^0,N_2^0)_L\sim (1,3,1/3),\\ \label{fer136}
\psi_{2 L}&=(N_3^0,E_1^-,E_2^-)_L\sim (1,3,-2/3),\quad E_{2L}^+,
\end{align}
\end{subequations}
with the following particle content: seven up type quarks, five down type quarks, ten charged lepton states and seven Weyl neutral lepton states.

\subsection{\label{sec:sec31}Weak currents}
The fermion currents for this particular Fermion structure are:
\subsubsection{Charged currents}
The interactions of the charged vector gauge boson fields with the spin 1/2 fermion fields are
\begin{eqnarray}\nonumber
H^{CC}&=&\frac{g}{\sqrt{2}}\sum_{i=1}^3\sum_\alpha\left\{\left[ W^+_\mu (\bar{\nu}_{\alpha L}\gamma^\mu \alpha_{L}^- + \bar{E}_{1L}^+\gamma^\mu N_{1L}^0 +\bar{N}_{3L}^0\gamma^\mu E_{1L}^- - \bar{u}_{i L}\gamma^\mu d_{iL}\right.\right. \\ \nonumber 
&& + \bar{u}_{4 L}\gamma^\mu d_{4L}) + K^+_\mu (\bar{\nu}_{\alpha L}\gamma^\mu E_{\alpha L}^- 
+\bar{E}_{1}^+\gamma^\mu N_{2L}^0 + \bar{N}_{3 L}^0\gamma^\mu E_{2L}^-\\ \nonumber 
&&- \bar{U}_{i L}\gamma^\mu d_{iL} + \bar{u}_{4 L}\gamma^\mu D_{L}) + K^0_\mu (\bar{\alpha}_{L}^-\gamma^\mu E_{\alpha L}^- +  \bar{N}_{1L}^0 \gamma^\mu N_{2L}^0 \\ 
&& \left.\left. + \bar{E}_{1L}^-\gamma^\mu E_{2L}^- - \bar{U}_{i L}\gamma^\mu u_{iL} + \bar{d}_{L}\gamma^\mu D_{4L})\right]\right\} + H.c., 
\end{eqnarray}
where $\alpha=e,\mu,\tau,\tau^\prime$ is a fourth family lepton index and $i=1,2,3$  is a three family quark index.

\subsubsection{Neutral currents}
The neutral currents $J_\mu(EM)$, $J_\mu(Z)$ and $J_\mu(Z')$, are associated with the Hamiltonian $H^0=eA^\mu J_\mu(EM)+\frac{g}{C_W}Z^\mu J_\mu(Z) + \frac{g'}{\sqrt{3}}Z'^\mu J_\mu(Z')$.

The vector-like electromagnetic current for this model is
\begin{eqnarray}\nonumber 
J_\mu(EM)&=&\frac{2}{3}\sum_{i=1}^3\left[(\bar{u}_{i}\gamma^\mu u_{i} + \bar{U}_{i}\gamma^\mu U_{i}+ \bar{u}_{4}\gamma^\mu u_{4}) -\frac{1}{3}(\bar{d}_{i}\gamma^\mu d_{i}+\bar{d}_{4}\gamma^\mu d_{4}+\bar{D}\gamma^\mu D)\right] \\ \nonumber
&& -\sum_\alpha(\bar{\alpha}^-\gamma^\mu \alpha^- + \bar{E}_{\alpha}^-\gamma^\mu E_{\alpha}^- +\bar{E}_{1}^-\gamma^\mu E_{1}^-+\bar{E}_{2}^-\gamma^\mu E_{2}^-)\\ 
&=& \sum_f q_f\bar{f}\gamma^\mu f,
\end{eqnarray}
where again the sum over $\alpha$ is for $\alpha=e,\mu,\tau,\tau^\prime$. The square root of the fine structure constant is proportional to $e\equiv g_3S_W=g_1C_W\sqrt{1-T^2_W/3}>0$, and $q_f$ is the electric charge of the fermion $f$ in units of $e$.

The left-handed currents for this model are 
\begin{eqnarray*}
J^\mu (Z) &=& J^{\mu}_L(Z)-S^2_WJ^\mu (EM),\\ 
J^\mu (Z') &=& T_WJ^\mu (EM)-J^{\mu}_L(Z'), 
\end{eqnarray*}
where
\begin{eqnarray}\nonumber 
J^{\mu}_L(Z) &=& \frac{1}{2}[\sum_\alpha\left(\bar{\nu}_{\alpha L}\gamma^\mu \nu_{\alpha L} - \bar{\alpha}_{L}^-\gamma^\mu \alpha_{L}^-\right)+\bar{E}_{1L}^+\gamma^\mu E_{1L}^+ -\bar{E}_{1L}^-\gamma^\mu E_{1L}^- + \bar{N}_{3L}^0\gamma^\mu N_{3L}^0- \bar{N}_{1L}^0\gamma^\mu N_{1L}^0\\ \nonumber && + \sum_{i=1}^3\left(\bar{u}_{iL}\gamma^\mu u_{iL} + \bar{u}_{4L}\gamma^\mu u_{4L} -\bar{d}_{iL}\gamma^\mu d_{iL} - \bar{d}_{4L}\gamma^\mu d_{4L}\right)\\ 
&=& \sum_f T_{3f}\bar{f}_L\gamma^\mu f_L,
\end{eqnarray}

\begin{eqnarray}\nonumber 
J^{\mu}_L(Z') &=& \sum_\alpha\sum_{i=1}^3[S_{2W}^{-1}(\bar{\nu}_{\alpha L}^0 \gamma^\mu \nu_{\alpha L}^0 +\bar{E}_{1L}^+\gamma^\mu E_{1L}^+ +\bar{N}_{3L}^0\gamma^\mu N_{3L}^0 -\bar{d}_{iL}\gamma^\mu d_{iL} + \bar{u}_{4L}\gamma^\mu u_{4L}) \\ \nonumber 
&& + T_{2W}^{-1}(\bar{\alpha}_{L}^-\gamma^\mu \alpha_{L}^- + \bar{N}_{1L}^0\gamma^\mu N_{1L}^0 + \bar{E}_{1L}^-\gamma^\mu E_{1l}^- - \bar{u}_{iL}\gamma^\mu u_{iL} + \bar{d}_{4L}\gamma^\mu d_{4L})\\ \nonumber
&& - T_{W}^{-1}(\bar{E}_{\alpha L}\gamma^\mu E_{\alpha L}^- +\bar{N}_{2L}\gamma^\mu N_{2L}+ 
\bar{E}_{2L}^-\gamma^\mu E_{2L}^-- \bar{U}_{iL}\gamma^\mu U_{iL}+\bar{D}_{L}\gamma^\mu D_{L})]\\ 
&=& \sum_fT_{9f}\bar{f}_L\gamma^\mu f_L,
\end{eqnarray}
where $S_{2W}=2S_WC_W$, $T_{2W}=S_{2W}/C_{2W}$, $C_{2W}=C_W^2-S_W^2$, $T_{3f}=Diag(1/2,-1/2,0)$ is the third component of the weak isospin acting on the representation 3 of $SU(3)_L$ (the negative when acting of $3^*$), and $T_{9f}=Diag(S_{2W}^{-1},$ $T_{2W}^{-1},-T_W^{-1})$ is a convenient 3x3 diagonal matrix acting on the representation 3 of $SU(3)_L$ (the negative when acting on $3^*$).

\section{\label{sec:sec4}Masses for Fermions}
In this section we calculate the most general mass matrices for the fermion fields of this particular model, produced by the 3 scalar Higgs fields in (\ref{higg}) and their respective VEV. We also set the notation to be used in the rest of the paper.
\subsection{\label{sec:sec41}Neutral Leptons}
The mass terms for the neutral Weyl states of this particular model are included in the following Yukawa Lagrangian:
\begin{equation}\label{lan13}
{\cal L}_n^{(13)}=\sum_{A=1,2}\left[\Phi_A^*\psi_{1L}\left(\sum_\alpha h^A_{\alpha}\psi_{\alpha L}+ h^A\psi_{2L}\right)\right]+h.c.,
\end{equation}
where again $\alpha=e,\mu,\tau,\tau^\prime$. Now, in the basis $(\nu_e^0,\nu_\mu^0,\nu_\tau^0,\nu_{\tau^\prime}^0,N_1^0,N_2^0,N_3^0)_L$ the former expression produces the following Hermitian mass matrix
\begin{equation}\label{man13}
M_N=\frac{1}{\sqrt{2}}\left(
\begin{array}{ccccccc}
0 & 0 & 0 & 0 & -M_e & m_e & 0 \\
0 & 0 & 0 & 0 & -M_\mu & m_\mu & 0 \\
0 & 0 & 0 & 0 & -M_\tau & m_\tau & 0 \\
0 & 0 & 0 & 0 & -M_{\tau^\prime} & m_{\tau^\prime} & 0 \\
-M_e & -M_\mu & -M_\tau & -M_{\tau^\prime} & 0 & 0 & -M \\
m_e & m_\mu & m_\tau & m_{\tau^\prime} & 0 & 0 & m \\
0 & 0 & 0 & 0 & -M & m & 0 \\
\end{array}\right),
\end{equation}
where the entries $M_\alpha=\sum_Ah_\alpha^AV_A,\;\; \alpha=e,\mu,\tau,\tau^\prime$ and $M=\sum_Ah^AV_A$ are Majorana mass terms proportional to the 3-3-1 mass scale $V_A,\; A=1,2$, and 
$m_\alpha=\sum_Ah_\alpha^Av_A,\;\; \alpha=e,\mu,\tau,\tau^\prime$ and $m=\sum_Ah^Av_A$ are
Dirac mass terms proportional to the SM mass scale $v_A,\; A=1,2$. This Hermitian mass matrix is a rank four matrix, with three eigenvalues equal to zero that we may identify as the three neutrinos, with the other four Weyl states pairing to produce Dirac masses at the scales $V_A$ and $v_A$ respectively.

\subsection{\label{sec:sec42}Charged leptons}
This model contains 10 charged leptons. The scalar Higgs fields in (\ref{higg}) couple to the charged 3-3-1 singlet leptons with the following Lagrangian
\begin{eqnarray}\label{lal13}\nonumber
{\cal L}_l^{(13)}&=&\sum_{A=1,2}\Phi_A\left\{h_{22}^A\psi_{2L}E_{2L}^+ 
+\sum_\alpha [h_{\alpha 2}^A\psi_{\alpha L}E_{2L}^+ \right. \\
&&+\left.\sum_{\beta^+_L}\left(h_{2\beta^+}^A\psi_{2L}\beta_L^+ 
+h_{\alpha\beta^+}^A\psi_{\alpha L}\beta_L^+\right)]\right\} + h.c.,
\end{eqnarray}
where $\alpha=e,\mu,\tau,\tau^\prime$ and $\beta_L^+=e_L^+,\mu_L^+, \tau_L^+,\tau^{\prime +}_L, E_{eL}^+,
E_{\mu L}^+,E_{\tau L}^+,E_{\tau^\prime L}^+$. For the most general case, the $10 \times 10$ mass matrix obtained is of rank ten.

\subsection{\label{sec:sec43}Neutrinos}
\begin{figure}
\includegraphics{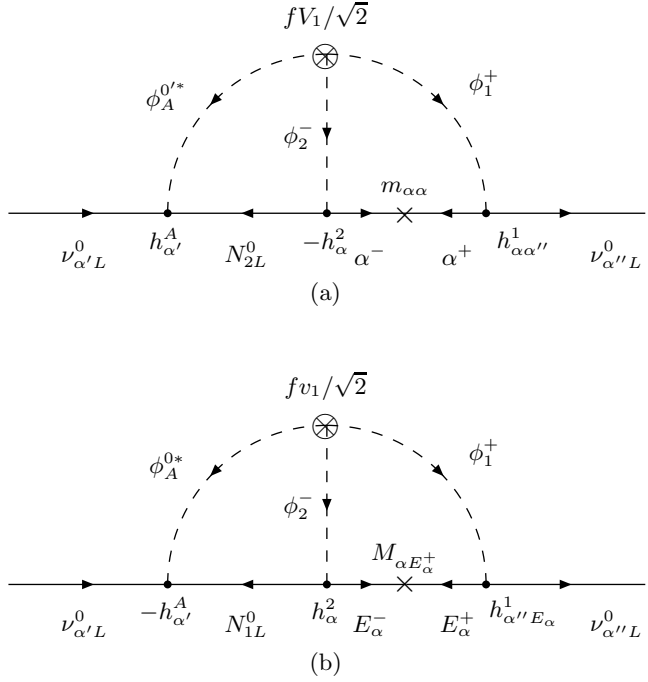}
\caption{\label{fig1}Two-loop Babu type diagrams contributing to the neutrino masses}
\end{figure}

The tree-level Hermitian $7\times 7$ mass matrix for the neutral spin 1/2 Weyl states in (\ref{man13}) has 3 eigenvalues equal to zero that we may identify with the 3 neutrinos in nature. Due to the richness of the model, those states pick up radiative Majorana masses when the quantum corrections are taken into account. As is inferred  from a careful study of the Yukawa Lagrangian in (\ref{lan13}) and (\ref{lal13}) we can draw two loop radiative diagrams Babu type~\cite{babu}, as depicted in Fig~(\ref{fig1}), which are contained in the model in a natural way, that is, without the inclusion of new ingredients. The upper vertex in the two diagrams comes from a term in the scalar potential of the form $f|\Phi_1^\dagger.\Phi_2|^2$.

\subsection{\label{sec:sec44}Up Quarks}
The Yukawa Lagrangian that the three Higgs scalar fields in Sec.~(\ref{sec:sec23}) produce for 
the seven up quark fields in model 13 is:
\begin{eqnarray}\label{lagu}
\mathcal{L}_u^{(13)}&=&\sum_{i=1}^{3} \sum_{A=1}^{2}\psi_{QL}^{iT} \phi^{*}_A C \left(\sum_{j=1}^{4}h_{iA}^{u_j} u_{L}^{jc}+\sum_{l=1}^3h_{iA}^{U_l}U_{L}^{lc}\right)\\ \nonumber 
&&+ \psi_{QL}^{4T} \phi_3 C \left(\sum_{j=1}^{4}h_{43}^{u_j} u_{L}^{jc}+
\sum_{l=1}^{3} h_{43}^{U_l} U_{L}^{lc}\right)+ h.c.,
\end{eqnarray}
where the $h's$ are Yukawa couplings and $C$ is the charge conjugation operator. For the most 
general case, the $7\times 7$ mass matriz obtained is of rank seven.


\subsection{\label{sec:sec45}Down quarks}
For the five Down quarks fields in model 13, the Yukawa Lagrangian that the three Higgs 
scalar fields and their VEV in Sec.~(\ref{sec:sec23}) produce, is:
\begin{eqnarray}\label{dqla}
\mathcal{L}_d^{(13)}&=&\sum_{i=1}^{3}\psi_{QL}^{iT} \phi_3^* C \left(\sum_{j=1}^{4}h_{i3}^{d_j} d_{L}^{jc}+h_{i3}^{D}D_{L}^c\right)\\ \nonumber 
&&+ \sum_{A=1}^2\psi_{QL}^{4T} \phi_A C \left(\sum_{j=1}^{4}h_{4A}^{d_j}d_{L}^{jc}+h_{4A}^{D} D_{L}^c\right)+ h.c.,
\end{eqnarray}
where the $h's$ are Yukawa couplings and $C$ is the charge conjugation operator. For the most 
general case, the $5\times 5$ mass matriz obtained is of rank five. 


The mass matrices obtained from (\ref{lagu}) and (\ref{dqla}) must be diagonalized in order to get the mass eigenstates which exist in nature, defining in this way a non unitary $7\times 5$ quark mixing matrix of the form 
\begin{eqnarray}\label{dmix}
V_{mix}^{ud}&\equiv& V_L^u{\cal P}V_L^{d\dag} \\ \nonumber 
&=&\left(\begin{array}{ccccc} 
V_{ud} & V_{us} & V_{ub} & V_{ub^\prime} &V_{ub^{\prime\prime}}  \\ 
V_{cd} & V_{cs} & V_{cb} & V_{cb^\prime} & V_{cb^{\prime\prime}} \\
V_{td} & V_{ts} & V_{tb} & V_{tb^\prime} & V_{tb^{\prime\prime}} \\
V_{t^\prime d} & V_{t^\prime s} & V_{t^\prime b} & V_{t^\prime b^\prime}& V_{t^\prime b^{\prime\prime}} \\
V_{t^{\prime\prime} d} & V_{t^{\prime\prime} s} & V_{t^{\prime\prime} b} & V_{t^{\prime\prime} b^\prime}& 
V_{t^{\prime\prime} b^{\prime\prime}} \\
V_{t^{\prime\prime\prime} d} &V_{t^{\prime\prime\prime} s} & V_{t^{\prime\prime\prime} b} & V_{t^{\prime\prime\prime} b^\prime} & 
V_{t^{\prime\prime\prime} b^{\prime\prime}} \\
V_{t^{iv\prime} d} &V_{t^{iv\prime} s} & V_{t^{iv\prime} b} & V_{t^{iv\prime} b^\prime} & 
V_{t^{iv\prime} b^{\prime\prime}} \\
\end{array}\right),
\end{eqnarray}
where $V_L^u$ and $V_L^d$ are $7\times 7$ and $5\times 5$ unitary matrices which diagonalize $M_UM_U^\dag$ and $M_DM_D^\dag$, with $M_U$ and $M_D$ the up and down quark mass matrices obtained from (\ref{lagu}) and (\ref{dqla}) repectively, and ${\cal P}$ is the projection matrix over the ordinary quark sector (in the weak basis, the exotic quarks transform as singlets under $SU(2)_L$ transformations, thus they do not couple with the $W^\pm$ Gauge Bosons). The transpose of this matrix is given by
\begin{equation}\label{calp}
{\cal P}^T=\left(\begin{array}{ccccccc} 
1 & 0 & 0 & 0 & 0 & 0 & 0 \\
0 & 1 & 0 & 0 & 0 & 0 & 0 \\
0 & 0 & 1 & 0 & 0 & 0 & 0 \\
0 & 0 & 0 & 0 & 0 & 0 & 0 \\ 
0 & 0 & 0 & 0 & 0 & 0 & 0 \\ 
               \end{array}\right).
\end{equation}
$V_{mix}^{ud}$ in (\ref{dmix}) defines the couplings of the physical quark states.
$(u,c,t,t^\prime,t^{\prime\prime},t^{\prime\prime\prime},t^{iv\prime})$ and $(d,s,b,b^\prime,b^{\prime\prime})$ with the charged current associated with the weak gauge boson $W^+$.

\section{\label{sec:sec5}Quark Mixing.}
In this section we are going to see how large the mixing between the known quarks in the first 3 families and the new ones can be, without violating current experimental limits. Two kinds of experimental constrains are going to be considered: the measured values of the known $3\times 3$ quark mixing matrix and current values and bounds for FCNC processes.

\subsection{\label{sec:sec51}The $3\times 3$ quark mixing matrix}
The masses and mixing of quarks in the SM come from Yukawa interaction terms with the Higgs condensate, which produces two $3\times 3$ quark mass matrices for the up and down quark sectors; matrices that must be diagonalized in order to identify the mass eigenstates. The unitary CKM quark mixing matrix ($V_{CKM}\equiv V^u_{3L}V^{d\dag}_{3L}$) couples the six physical quarks to the charged weak gauge boson $W^+$, where $V_{3L}^u$ and $V_{3L}^d$ are now the diagonalizing unitary $3\times 3$ matrices of the SM up and down quark sectors respectively.

The unitary matrix $V_{CKM}$ has been parameterized in the literature in several different ways, but the most important fact related with this matrix is that most of its entries have been measured with high accuracy, with the following experimental limits~\cite{pdg}:
\begin{widetext}
\begin{equation}\label{maexp}
 V_{exp}=
\left(\begin{array}{ccc}
0$.$970\leq  |V_{ud}|\leq 0$.$976 & 0$.$223\leq |V_{us}|\leq 0$.$228 & 0$.$003\leq |V_{ub}|\leq0$.$006\\
0$.$219\leq |V_{cd}| \leq 0$.$241 & 0$.$90\leq |V_{cs}| \leq 1$.$0 & 0$.$037 \leq |V_{cb}| \leq 0$.$045\\
0$.$006\leq |V_{td}| \leq 0$.$008 & 0$.$034\leq |V_{ts}| \leq 0$.$044 & |V_{tb}| \geq0$.$78
\end{array}\right).
\end{equation}
\end{widetext}

The numbers quoted in matrix (\ref{maexp}), which are measured at the Fermi scale $(\mu~\approx M_Z)$~\cite{koide}, are generous in the sense that they are related to the direct experimental measured values, at 90\% confidence level, with the largest uncertainties taken into account, without bounding the numbers to the orthonormal  constrains on the rows and columns of a $3\times 3$ unitary matrix. In this way we leave the largest room available for possible new physics, respecting the well measured values in $V_{exp}$.

\subsection{\label{sec:sec52}FCNC}
The unitary character of the SM mixing matrix $V_{CKM}$ implies flavor diagonal couplings of all the neutral bosons of the SM (such as Z boson, Higgs boson, gluons and photon) to a pair of quarks, giving as a consequence that no FCNC are present at tree level. At one-loop level, the charged currents generate FCNC transitions via penguin and box diagrams~\cite{jf}, but they are highly suppressed by the GIM mechanism~\cite{sg}. For example, FCNC processes in the charm  sector $(c\rightarrow u\gamma)$ were calculated in the context of the SM in Ref.~\cite{GB}, giving a branching ratio suppressed by 15 orders of magnitude.

To date, the following direct FCNC branching ratios and bounds have been measured in several experiments:
\begin{itemize}
\item $\mathcal{B}r[b\rightarrow s\gamma]=(3.52\pm 0.24)\times 10^{-4}$ ~\cite{hfag}
\item $\mathcal{B}r[B\rightarrow K^*l^+l^-]= (1.68\pm 0.86)\times 10^{-6}$ ~\cite{BB},
\item $\mathcal{B}r[s\rightarrow d\gamma(dl^+l^-)]<10^{-8}$ ~\cite{sp}
\item $\mathcal{B}r[c\rightarrow ul^+l^-]<4\times 10^{-6}$ ~\cite{vm}
\item $\mathcal{B}r[b\rightarrow sl^+l^-,dl^+l^-]<5\times 10^{-7}$ ~\cite{VM},
\end{itemize}
with $l=e,\mu$. In our study, these ratios and bounds are also going to be respected. Important to mention here that the SM next to next to leading order calculation for $\mathcal{B}r[b\rightarrow s\gamma]$ is $(3.60\pm 0.30)\times 10^{-4}$~\cite{gambi}, already in agreement with the measured value, which   constitutes a very sensitive prove of new physics.

\subsection{\label{sec:sec53}Numerical Analysis.}
The numerical analysis which follows aims to set upper bounds on the fourth rows and fourth columns of $M_U$   and   $M_D$, the quark mass matrices obtained from (\ref{lagu}) and (\ref{dqla}) using as phenomenology constrains the values of the matrix $V_{exp}$ in (\ref{maexp}) and the experimental measured values and bounds for FCNC quoted in Sec.~(\ref{sec:sec52}). Since we want to consider just the maximal mixing of the ordinary quarks with the exotic ones, our approach consists of looking only for a mixing with the fourth family up and down quarks, under the assumption that the simultaneous mixing with the other exotic quarks is always possible, but not maximal. So, in our numerical analysis we will set zero  everywhere in the fifth row and column of $M_D$ except for the diagonal entry, doing the same for the  fifth, sixth and seven rows and columns of $M_U$.

In the analysis we assume that $v_1=v_2=v_3\equiv v=82$ GeV, value  supported by the result $M_W^2=g_3^2(v_1^2+v_2^2+v_3^2)/2$~\cite{vl}  with $g_3$ the gauge coupling constant of $SU(3)_L$ (that is equal to $g_2$, the gauge coupling constant of $SU(2)_L$ in the SM), and also we use $V=1$ TeV, the 3-3-1 mass scale which fixes the mass values for all the new fermions of the different models.

\subsection{\label{sec:sec54}The $7\times 5$ mixing matrix}
In this section we are going to set limits to some values of the mixing matrix $V_{mix}^{ud}$ in (\ref{dmix}). For this purpose we have designed a software which start by fixing a value $V_{tb}\sim 0.78$, the smallest permited by $V_{exp}$ in (\ref{maexp}). According to our analysis this allows for maximal mixing of the top quark with the other ones.

Notice also that, although the mass matrices $M_U$ and $M_D$ in Eqs.~(21) and (22) are not symmetric due to the existence of the new mass scale $V_A$ present in all the 3-3-1 models, they can always be made the product of a unitary matrix times a Hermitian one, with the unitary matrix reabsorbed in a new definition of the right-handed quark fields, something which does not affect the physics of 3-3-1 models, because the right-handed fields are singlets under $SU(3)_L$. So, for our numerical analysis we will look only for symmetric up and down quark mass matrices.

The systematic random numerical analysis using MATHEMATICA subroutines, throws as a result that the largest mixing allow of the known quarks with the exotic ones, without violating the measured values of the mixing matrix (\ref{maexp}), or the experimental values and  bounds for FCNC, occurs for the following set of mass matrices:

\begin{equation}\label{upte4}
M^u_7 =\left(\begin{array}{ccccccc} 
0.00047 & 0.02812 & 0 & 0 & 0 & 0 & 0\\ 
0.02812 & 0.580   & 0 & 0 & 0 & 0 & 0 \\
0 &  0 & 171.7  & 0 & 0 & 0 & 0 \\
0 & 0 & 0 & m_{t^\prime} & 0 & 0 & 0 \\
0 & 0 & 0 & 0 & m_{t^{\prime\prime}} & 0 & 0 \\
0 & 0 & 0 & 0 & 0 & m_{t^{\prime\prime\prime}} & 0 \\
0 & 0 & 0 & 0 & 0 & 0 & m_{t^{iv\prime}}\\
\end{array}\right)
\end{equation}

\begin{equation}\label{dowte5}
M^d_5 =\left(\begin{array}{ccccc} 
0.018 & -0.4288 & -2.63 & -3.41 & 0\\ 
-0.4288 & 9.316 & 57.608 & 75.98 & 0 \\
-2.63 & 57.608 & 361.8 & 472.4 & 0 \\
-3.41 & 75.98 & 472.4 & 624.5 & 0 \\
0 & 0 & 0 & 0 & m_{b^{\prime\prime}} \\
\end{array}\right),
\end{equation}
which for $m_{t^\prime}=m_{t^{\prime\prime}} =m_{t^{\prime\prime\prime}} =m_{t^{iv\prime}} =m_{b^{\prime\prime}}=1000$ GeV, reproduce the following set of eigenvalues at the weak scale (in units of GeV)
\[m_t=171.8,\; m_c=0.582, \; m_u=1\times 10^{-3}\]
\[m_b=2.83,\; m_s=0.069, \; m_d=3.4\times 10^{-3};\]
\[m_{b^\prime}=993, \]
numbers to be compared with the values quoted in the second paper in Ref.~\cite{koide}.

Diagonalizing the former mass matrices produce the following non-unitary $7\times 5$ mixing matrix.
\begin{equation}\label{mix45}
V_{mix}^{7\times 5} =\left(\begin{array}{ccccc} 
0.976 & 0.224 & 0.006 & 0.009 & 0 \\ 
0.220 & -0.970 & 0.038 & 0.096 & 0 \\
0.006 &  0.035  & 0.798 & 0.602 & 0 \\
0 & 0 & 0 & 0 & 0  \\
0 & 0 & 0 & 0 & 0  \\
0 & 0 & 0 & 0 & 0  \\
0 & 0 & 0 & 0 & 0  \\
\end{array}\right),
\end{equation}
numbers to be compared with the experimental limits in (\ref{maexp}).


\subsection{\label{sec:sec55} New FCNC processes}
Next, we are going to evaluate the new penguin contributions to the FCNC processes coming from the nonunitary character of 
$V^{7\times 5}_{mix}$ in Eq.~(\ref{mix45}).

\subsubsection{The bottom sector}
Let us evaluate first the electromagnetic penguin contribution to ${\cal B}r^{t}(b\rightarrow s\gamma)$ coming from the $t$ quark, calculated with the expectator model, scaled to the semileptonic decay $b\rightarrow q_il\nu_l,\; q_i=c,u$, and without including QCD corrections (which are small for the $b$ sector~\cite{jf}). This value is calculated to be~\cite{GB}

\begin{equation}\label{fincal}
{\cal B}r^{t}(b\rightarrow s\gamma)\approx\frac{3\alpha}{2\pi}
\frac{|V_{tb}^*V_{ts}F^Q(x^{t})|^2}{[f(x_c)|V_{cb}|^2+f(x_u)|V_{ub}|^2]}B_{B\rightarrow Xl\nu_l},
\end{equation}
where $\alpha$ is the fine structure constant, $B_{B\rightarrow Xl\nu_l}\approx 0.1$ is the branching ratio for semileptonic $b$ meson decays taken from Ref.~\cite{pdg}, $x^{t}=(m_{t}/M_W)^2$, $x_c=m_c/m_b$ and $x_u=m_u/m_b$. $F^Q(x)$ is the contribution of the internal heavy quark line to the electromagnetic penguin given by

\begin{eqnarray*}
F^Q(x)&=&Q\left[\frac{x^3-5x^2-2x}{4(x-1)^3}+\frac{3x^2\ln{x}}{2(x-1)^4}\right] \\
&+&\frac{2x^3+5x^2-x}{4(x-1)^3}-\frac{3x^3\ln{x}}{2(x-1)^4},
\end{eqnarray*}
where $Q=2/3$ for $t$ in the quark propagator [$Q=-1/3$ and $x=x^{b^\prime}=(m_{b^\prime}/M_W)^2$ when $b^\prime$ propagates] and $f(x_i)$ is the usual phase space factor in semileptonic meson decay, given by~\cite{jf}

\[f(x)=1-8x^2+8x^6-x^8-24x^4\ln{x}.\]

For the numerical evaluations of ${\cal B}r^{t}(b\rightarrow s\gamma)$, let us use the values $\alpha(1GeV)=1/135$, $m_{t}=385$ GeV, $m_c=1.25$ GeV, $m_b=6.0$ GeV and $m_u=2.6$ MeV, which are the mass values at 1 GeV~\cite{koide}. Using these numbers we obtain: $F^{2/3}(x^{t})\approx 0.544$, $f(x_c)\approx 0.724$ and $f(x_u)\approx 1$. Plug in the numbers in Eq.~(\ref{fincal}) and using the values for $V_{mix}^{7\times 5}$ in equation (\ref{mix45}) for the couplings of the physical quark states, we get

\[{\cal B}r^{t}(b\rightarrow s\gamma)\approx 6\times 10^{-5},\]
close to the SM calculation as it should be, since this process does not receive a contribution from the exotic quarks.

The former analysis can be used also to estimate the branching ratios for the rare gluon penguin decay $b\longrightarrow sg$, where $g$ stands for the gluon field. The results is
\begin{eqnarray*}
{\cal B}r^{t}(b\rightarrow sg)&=&\frac{\alpha_s(1GeV)}{\alpha(1GeV)}{\cal B}r^{t}(b\rightarrow s\gamma)\\ 
&\approx&13{\cal B}r^{t}(b\rightarrow s\gamma)\approx 7.8\times 10^{-4},
\end{eqnarray*}
a process difficult to measure due to the hadronization of the gluon field $g$. (This last process is of the same order of magnitude of the virtual weak penguin bottom process $b\longrightarrow sZ$).

A similar analysis shows that 
\[{\cal B}r^{t}(b\rightarrow d\gamma)=\frac{|V_{td}|^2}{|V_{ts}|^2}{\cal B}r^{t}(b\rightarrow s\gamma)\approx 2.2\times 10^{-6},\]
which is safe and in agreement with the bounds quoted in Section (\ref{sec:sec42}).

\subsubsection{The strange sector}
In a similar way we can evaluate ${\cal B}r^{t}(s\rightarrow d\gamma)$ scaled to the semileptonic decay $s\rightarrow ul\nu_l$, which is given now by 
\begin{equation}\label{fincs}
{\cal B}r^{t}(s\rightarrow d\gamma)\approx\frac{3\alpha}{2\pi}
\frac{|V_{ts}^*V_{td}F^{2/3}(x^{t})|^2}{f(x^\prime_u)|V_{us}|^2}B_{K\rightarrow \pi l\nu_l}.
\end{equation}

With $x^\prime_u=m_u/m_s,\; m_s$(1GeV)=111 MeV, and $B_{K\rightarrow \pi l\nu_l}\approx 5\times 10^{-2}$ taken from Ref.~\cite{pdg}, we get 
\[{\cal B}r^{t}(s\rightarrow d\gamma)\approx 3.8\times 10^{-11},\]
in agreement with the experimental bound quoted in Section (\ref{sec:sec42}).

\subsubsection{The charm sector}
Now let us evaluate ${\cal B}r^{b^\prime}(c\rightarrow u\gamma)$ scaled to the semileptonic decay $c\rightarrow q_jl\nu_l$, where $q_j=s,d$. The branching ratio is

\begin{equation}\label{finccu}
\frac{{\cal B}r^{b^\prime}(c\rightarrow u\gamma)}{B_{D\rightarrow X_s l\nu_l}}\approx\frac{3\alpha}{2\pi}
\frac{|(V_{cb^\prime}^*V_{ub^\prime}) F^{-1/3}(x^{b^\prime})|^2}{[f(x_s)|V_{cs}|^2+f(x_d)|V_{cd}|^2]},
\end{equation}
where $x_s=m_s/m_c,\; x_d=m_d/m_c$. With $B_{D\rightarrow X_s l\nu_l}\approx 0.2$ taken from Ref.~\cite{pdg}, $F^{-1/3}(x^{b^\prime})\approx 0.3856$, $f(x_s)\approx 0.94$ for $m_s=111$ MeV and $f(x_d)\approx 1$, for $m_d=5.6$ MeV, we get 
\[{\cal B}r^{b^\prime}(c\rightarrow u\gamma)\approx 8.36\times 10^{-8},\]
seven orders of magnitude larger than the SM prediction~\cite{GB}, but still unobservable small. Of course, the quantum QCD corrections for this decay could be quite large (see the second paper in Ref.~\cite{GB}).

\subsubsection{The top sector}
We proceed this analysis with the study of the FCNC for the top quark in the context of the four family 3-3-1 model in consideration here. As we are about to see, some of the predictions are ready to be tested at the Large Hadron Collider (LHC).

In the SM, the one-loop induced FCNC for the top quark have a strong GIM suppression, resulting in negligible branching ratios for top FCNC decays. The SM values predicted are~\cite{saav}: ${\cal B}r^{SM}(t\rightarrow c\gamma)\approx 4.6\times 10^{-14}$, and 
${\cal B}r^{SM}(t\rightarrow cg)\approx 4.6\times 10^{-12}$.

The new FCNC ${\cal B}r^{b^\prime}(t\rightarrow c\gamma)$ and ${\cal B}r^{b^\prime}(t\rightarrow u\gamma)$ predicted for the top quark in the context of the 3-3-1 model under consideration here, scaled to the semileptonic decay $t\rightarrow q_k l\nu_l,\; q_k=b,s,d$; are given by 

\begin{equation}\label{finct}
\frac{{\cal B}r^{b^\prime}(t\rightarrow c\gamma)}{B_{T\rightarrow Xl\nu_l}}\approx\frac{3\alpha}{2\pi}
\frac{|(V_{tb^\prime}^*V_{cb^\prime}) F^{-1/3}(x^{b^\prime})|^2}{[f(x_b)|V_{tb}|^2+f(x_s)|V_{ts}|^2]}
\end{equation}
which we evaluate at the $m_t=163$ GeV, the pole mass scale for the top quark, which gives

\[{\cal B}r^{b^\prime}(t\rightarrow c\gamma)\approx 2.76\times 10^{-6}B_{T\rightarrow Xl\nu_l},\]
which is large as far as the semileptonic branching ratio $B_{T\rightarrow Xl\nu_l}$ measured for the top quark gets comparatively large, and much larger than $10^{-14}$, the SM prediction.

From the former analysis we can get 
\[{\cal B}r^{b^\prime}(t\rightarrow cZ)=\frac{4\pi}{\sin(2\theta)}{\cal B}r^{b^\prime}(t\rightarrow c\gamma)\approx 40 
{\cal B}r^{b^\prime}(t\rightarrow c\gamma),\]
two orders of magnitude larger than ${\cal B}r^{b^\prime}(t\rightarrow c\gamma)$, a value not far from the LHC capability, with a similar conclusion for the branching ${\cal B}r^{b^\prime}(t\rightarrow cg)$, where $g$ stands for the gluon field.

Finally we find 
\begin{eqnarray*}
{\cal B}r^{b^\prime}(t\rightarrow u\gamma)&\approx&\frac{|V_{ub^\prime}|^2}{|V_{cb^\prime}|^2}{\cal B}r^{b^\prime}(t\rightarrow c\gamma)\\
&\approx& 2.43\times 10^{-8} B_{T\rightarrow Xl\nu_l}.
\end{eqnarray*}

\section {\label{sec:sec6} Conclusions}
In the present work we classified all the possible 3-3-1 four family models which do not contain exotic electric charges neither fermion vector-like representations (including right-handed neutrino singlets). A total of 13 different fermion structures were found, out of which only two are realistic in the sense that they can bear 3 and only 3 light neutrinos.

Contrary to the minimal 3-3-1 model of Pisano-Pleitez and Frampton~\cite{pf} and some of its trivial extensions which, due to the particular way the anomalies are cancelled, accept only models for a number of families equal to a multiple of 3 (and just 3 if the QCD asymptotic freedom is recalled), the models without exotic electric charges can hold for as many families as wished. So, if a fourth family is found for example in the LHC experiments, the minimal 3-3-1 model will be ruled-out, but not the models we have presented here.

Also, in the paper we did some phenomenology for what we believe is the simplest realistic 3-3-1 four family model. In particular, for the model under study we searched for the largest mixing between ordinary and exotic quarks without violating current experimental constrains in the quark mixing matrix and in the values and bounds measured for FCNC processes.

Some of our conclusions may be relevant for the forthcoming Tevatron and LHC results, which may find evidence for a fourth family and also measure with high accuracy the value of $V_{tb}$; in particular, a value in the range $0.8\leq V_{tb}\leq 0.9$ can lead to strong predictions of rare top decays such as $t\rightarrow cZ$, with a branching ratio of the order of $10^{-5}$, perfectly reachable at the LHC~\cite{carv}.

To conclude, let us spell out our main conclusion: ``the existence of a fourth family does not rule out the $SU(3)_c\otimes SU(3)_L\otimes U(1)_X$ local gauge structure''.\\

\textbf{Aknowdledgements:} We aknowledge partial finantial support from the "sostenibilidad'' program of the Universidad de Antioquia, Colombia.


\end{document}